\documentclass[twocolumn,epsfig,aps,prb]{revtex4}

\begin{document}

\title{ Experimental observation of speckle instability in nonlinear disordered media.}

\author{Igor I. Smolyaninov, Ali Gungor $^*$, and Christopher C. Davis}
\address{Department of Electrical and Computer Engineering, University of Maryland, College Park, MD 20742, USA}

\address{$^*$ Department of Engineering, Bahcesehir University, Istanbul, 34538, Turkey}

\date{\today}

\begin{abstract}
Temporal fluctuations of the speckle pattern formed upon backscattering of a laser beam from an interface between gold and nonlinear polymer film have been observed as a function of optical power. The instability can be explained by coupling of laser light to surface plasmons and other guided modes, which experience multiple scattering while propagating in the film along the interface. The speckle pattern produced in this process is extremely sensitive to fluctuations of the scattering potential near the interface. 
\end{abstract}


\maketitle

Propagation of coherent waves in nonlinear disordered media is the topic of considerable current interest. Many experimental situations, such as light propagation in weakly disordered Kerr media \cite{1}, the theory of hydrodynamic turbulence \cite{2}, the theory of turbulent plasma \cite{3}, the mesoscopic theory of interacting electrons in disordered metals \cite{4}, etc. can be described by a nonlinear Schrodinger equation in which a field $\phi(r)$ propagation in an elastically scattering medium is affected by both the medium potential $u(r)$, and by the nonlinear self-interaction potential $\beta n(r)=\beta|\phi(r)|^2$. While in the linear $\beta=0$ limit the field distribution exhibits speckle (which means that $n(r)$ becomes random, sample specific function of coordinates), it has been shown very recently \cite{4,5} that in the nonlinear case the speckle pattern exhibits instability, and hence temporal fluctuations. The striking feature of this theoretical result is that the instability occurs even for very small nonlinearities. This conclusion has important implications for wide variety of physical systems described above. For example, this phenomenon may have very important consequences for random lasers \cite{6}. Unfortunately, despite very strong theoretical interest, there was no experimental observation of speckle instability in nonlinear media.

In this paper we report the first to our knowledge experimental observation of this effect. We present the experimental study of temporal fluctuations of the speckle pattern formed upon backscattering of a laser beam from an interface between gold and nonlinear polydiacetylene film. The speckle fluctuations were studied as a function of optical power. The instability of the speckle pattern can be explained by coupling of laser light to surface plasmon polaritons (SPP) \cite{7} and other guided modes, which experience multiple scattering over their propagation in the film along the interface. The speckle pattern produced in this process is extremely sensitive to fluctuations of the scattering potential near the interface. 

Microscopic images of the poly-3-butoxy-carbonyl-methyl-urethane (poly-3BCMU) polydiacetylene film on gold film substrate used in this study are shown in Fig.1(a,b). This poly-3BCMU film was spin-coated onto the gold film surface from the solution in chloroform. Such 3BCMU polydiacetylene films have one of the largest known fast nonresonant optical $\chi^{(3)}$ nonlinearities \cite{8} (according to ref.\cite{8} various components of the $\chi^{(3)}$ tensor of the poly-3BCMU film have magnitudes of the order of $10^{-10}$ esu). The images in Fig.1 were obtained using cross-polarized detection, so that numerous orientational domains in the poly-3BCMU film are clearly visible. Fourier transformations of these images shown in (c,d) indicate the presence of large number of various random gratings, which are formed on the gold film surface by the periodic arrangements of the orientational domains. For example, the periodicity of such a grating visible inside the box in Fig.1(b) was measured to be $1.05 \mu m$. Formation of orientational gratings in various molecular films on gold surface is a rather usual effect (see for example ref.\cite{9}) Dielectric gratings on gold surface are known to cause efficient coupling of external illumination to surface plasmon polaritons \cite{7}. They will also efficiently couple light to other guided electromagnetic modes, which may propagate inside the poly-3BCMU film. Fig.1 also shows images of a $PbZr_xTi_{1-x}O_3$ (PZT) and a paper sample, which were used in our study as control samples.

Our experimental setup is shown in Fig.2. The speckle pattern has been measured in the backscattering geometry as a function of time and light intensity using a fast camera. The incident single-mode Gaussian laser beam (0.7 mm beam diameter) from a 21 mW He-Ne laser operated at 633 nm has been sent onto the sample surface at a small angle. According to numerous recent experimental and theoretical studies (see for example refs.\cite{10,11}) scattering of the incident light by the surface roughness of the gold film causes excitation of SPPs, which propagate in various directions along the metal surface, backscatter by the surface defects, and give rice to the enhanced backscattering in the direction opposite to the direction of incoming illumination. In the case of poly-3BCMU film on the gold film surface the coupling of light to SPP is also facilitated by various grating-like domain arrangements visible in Fig.1(a-d).

The characteristic time dependencies of the signals measured in some given pixel within the speckle pattern in the inset in Fig.2 are presented in Fig.3(a) for two different intensities of the incident laser beam. The speckle signal measured at higher power exhibits considerable noise compared to the signal measured at two times lower power on the same sample, and compared to control measurements performed at the same power levels on PZT (Fig.3(e)) and paper (Fig.3(c)) samples. This is also apparent from the Fourier spectra of these signals, which are also shown in Fig.3. Comparison of data obtained for gold-poly-3BCMU interface and for the paper surface allows us to exclude thermal and mechanical origins of the observed effect. The measurements performed on PZT sample allow us to compare the behavior of our thin film poly-3BCMU-on-gold sample with the properties of speckles produced by a bulk polycrystalline nonlinear optical material. Optical properties of PZT are close to the properties of such classic photorefractive materials as $BaTiO_3$ and $SrTiO_3$, which were studied in great detail. These materials have rather large $\chi^{(2)}$ and $\chi^{(3)}$ nonlinearities. For example, $\chi^{(3)}\approx 3\times 10^{-12}$ esu measured at 1064 nm is reported for $SrTiO_3$ \cite{12}. However, only in the $\sim 0.1$ Hz frequency range there is a noticeable difference in the noise spectrum obtained on speckles produced by the PZT sample at lower and higher optical powers. These slow speckle variations are consistent with the slow phase conjugation and holographic recording times observed in photorefractive materials at a comparable optical power (see for example ref. \cite{13}).  

In addition to temporal variations, the speckle pattern observed on the poly-3BCMU sample exhibited considerable spatial variations as a function of incident laser power. This is very apparent from the spatial self-correlation functions of the speckle patterns obtained at different optical powers, which are shown in Fig.4. The self-correlation functions obtained at lower power in (a) and (b) typically exhibited a three-lobe structure, which was probably caused by the domain gratings in the poly-3BCMU film (see Figs.1(a-d)). This explanation was confirmed in the experiment in which the sample had been rotated (which probably also caused a lateral displacement of the illuminated area on the sample surface). The self correlation function shown in Fig.4(c) was obtained after Fig.4(b) when the incident laser power had been increased by roughly a factor of two. A more complicated structure of the self-correlation function measured at higher power is consistent with theoretical results of refs.\cite{4,5}, which predict that at higher light intensity the speckle pattern experiences fast jumps back and forth between various different solutions of the nonlinear Shrodinger equation. Thus, our experimental results clearly demonstrate considerable temporal and spatial variations of the speckle pattern in nonlinear optical medium, which start to occur at some threshold optical power. It is interesting that the difference between noise power spectra in Fig.3(b) plotted in Fig.5 using the log-log coordinates exhibits classical 1/f behavior, which was previously observed in the noise spectra of many other mesoscopic systems.  

Qualitatively, the behavior of the speckle pattern produced by the laser reflection from the gold-poly-3BCMU interface exhibits all the features predicted theoretically in refs. \cite{4,5}. However, in order to achieve quantitative agreement with the numerical results of refs.\cite{4,5} it is absolutely essential to assume efficient coupling of the incident light to surface plasmon polaritons and other guided electromagnetic modes, which may propagate in the polymer film along the gold-polymer interface. According to eq.(12) from ref.\cite{5} the onset of speckle pattern instability occurs at

\begin{equation}
\label{eq1}
\Delta n^2(\frac{L}{l})^2(kl+\frac{L}{l})\approx 1 ,
\end{equation}

where $\Delta n=n_2I$ is the characteristic change in the refractive index of the nonlinear material due to external illumination, $L$ is the absorption length, $l$ is the mean free path between the scattering events, and $k$ is the wave vector. Assuming macroscopic $L\sim 1$ mm, Skipetrov and Maynard obtained $\Delta n\sim 10^{-6}-10^{-5}$ for the speckle instability threshold (notice though that in the absence of losses the speckle pattern is unstable even at infinitely small nonlinearities). Even though very small, such refractive index changes are much larger than $\Delta n\sim 10^{-10}$ which may be expected in the bulk nonlinear optical material with $\chi^{(3)}\sim 10^{-10}$ esu at the characteristic optical intensities $\sim 5$ W/cm$^2$ of the unfocused He-Ne laser used in our experiments. Thus, we must explain roughly four orders of magnitude quantitative discrepancy in the $\Delta n$ value, which corresponds to the onset of speckle pattern instability. On the other hand, efficient coupling of incident laser light to surface plasmons and other guided modes in the polymer film, which is facilitated by the domain gratings shown in Fig.1, may help us to overcome this inconsistency. It is well-known that under the phase-matching conditions established either due to use of Kretschmann geometry, or due to grating-assisted coupling, up to 90 percent of the incident light may be coupled to SPPs \cite{7}, while coupling to guided modes in the polymer film may be no less efficient. Assuming $\sim 10 \mu m$ polymer film thickness and close to 100 percent coupling, the electric field in the guided wave should be enhanced by a factor of 10 compared to the incident laser beam. Coupling of incident light to SPPs would result in the electric field enhancement factor of the order of 100 (due to the fact that the SPP field is localized within about 200 nm from the metal-dielectric interface). Such field enhancement factors are indeed typically observed in the experiments with surface plasmon polaritons \cite{7}. In addition, local values of the electric field in the surface electromagnetic wave may be further enhanced by surface roughness. An electric field enhancement by a factor of 100 is already sufficient to reach the $\Delta n\sim 10^{-6}-10^{-5}$ level of nonlinearities, which should induce the speckle instability according to refs.\cite{4,5}.      

Additional factor, which may play an important role in inducing the speckle instability in the poly-3BCMU-on-gold sample is the presence of refractive index fluctuations near the gold-polymer interface due to the quantum highway mirage effect \cite{14}. The theoretical results in refs.\cite{4,5} were obtained assuming that an infinitely weak noise is driving the speckle pattern jumps between different solutions of the nonlinear Shrodinger equation (the number of such solutions may be infinite, as shown in \cite{4}). According to ref.\cite{14} the fluctuations of the refractive index near the metal-dielectric interface are very strong. This factor may contribute to our observations of the speckle pattern fluctuations at much lower nonlinearities than the ones predicted in refs.\cite{4,5}. 

In conclusion, we have observed temporal fluctuations of the speckle pattern formed upon backscattering of a laser beam from an interface between gold and nonlinear polymer film. This instability can be explained by coupling of laser light to surface plasmons and other guided modes, which experience multiple scattering while propagating in the film along the interface. The speckle pattern produced in this process is extremely sensitive to fluctuations of the scattering potential near the interface, which may be induced by the quantum highway mirage effect. 

This work has been supported in part by the NSF grants ECS-0304046, CCF-0508213, and ECS-0508275.

\begin{figure}
\begin{center}
\end{center}
\caption{ (a) $350\times 470 \mu m^2$ and (b) $70\times 94 \mu m^2$  microscopic images of the poly-3-butoxy-carbonyl-methyl-urethane (poly-3BCMU) polydiacetylene film on gold film substrate (obtained using cross-polarized detection) visualize large number of orientational domains in the film. Fourier transformations of these images shown in (c) and (d), respectively, demonstrate various periodicities in the domain arrangement. The domain periodicity within the box indicated in (b) and shown by the arrow in (d) is $1.05 \mu m$. (e) and (f) show the $70\times 94 \mu m^2$ images of a $PbZr_xTi_{1-x}O_3$ (PZT) and paper samples also used in this study.}
\end{figure}

\begin{figure}
\begin{center}
\end{center}
\caption{Schematic view of our experimental setup. The spatial distribution of speckle (see the inset) in the backscattering geometry has been measured as a function of time using a fast camera.}
\end{figure}

\begin{figure}
\begin{center}
\end{center}
\caption{ (a) Time dependencies of the signals in a given pixel within the speckle pattern in the inset in Fig.2 measured for two different intensities of the incident laser beam. The power attenuation factor is shown near each curve. (b) Fourier spectra of the time dependencies in (a). Two other sets of plots represent data of control measurements performed under similar conditions on speckles produced by paper (c,d) and PZT (e,f) samples. }
\end{figure}

\begin{figure}
\begin{center}
\end{center}
\caption{Spatial self-correlation functions of the speckle patterns obtained from the gold-poly-3BCMU interface. The power attenuation factor is shown in each image. The self-correlation function in (b) was obtained after (a) as a result of sample rotation. From (b) to (c) the power of the incident laser beam was increased by roughly a factor of two. }
\end{figure}

\begin{figure}
\begin{center}
\end{center}
\caption{The difference between noise power spectra in Fig.3(b) plotted using the log-log coordinates exhibits classical 1/f behavior. }
\end{figure}


\begin{references}

\bibitem{1} L. Landau and E. Lifshitz, \textit{Electrodynamics of Continuous Media} (Pergamon, Oxford, 1968).

\bibitem{2} V.E. Zakharov, V.S. Lvov, and G. Falkovich, \textit{Kolmogorov Spectra of Turbulence} (Springer, Berlin, 1992).

\bibitem{3} B.B. Kadomzev, \textit{Collective Phenomena in Plasma} (Nauka, Moscow, 1976).

\bibitem{4} B. Spivak and A. Zyuzin, Phys.Rev.Lett. 84, 1970 (2000).

\bibitem{5} S.E. Skipetrov and R. Maynard, Phys.Rev.Lett. 85, 736 (2000).

\bibitem{6} D.S. Wiersma, M.P. van Albada, and A. Lagendijk, Nature 373, 203 (1995).

\bibitem{7} A.V. Zayats and I.I. Smolyaninov, J. Opt. A: Pure Appl. Opt. 5, S16 (2003).

\bibitem{8} K. Yang, W. Kim, A. Jain, J. Kumar, S. Tripathy, Optics Comm. 164, 203 (1999).

\bibitem{9} I.I. Smolyaninov, R. Coratger, F. Ajustron and J. Beauvillain, Phys.Letters A, 181, 251 (1993).


\bibitem{10} A.R. McGurn, A.A. Maradudin, and V. Celli, Phys.Rev.B 31, 4866 (1985).

\bibitem{11} C.S. West and K.A. O'Donnell, Phys.Rev.B 59, 2393 (1999).

\bibitem{12} M.J. Weber, \textit{Handbook of Optical Materials} (CRC Press, Boca Raton, 2003).

\bibitem{13} S.I. Bozhevolnyi, O. Keller and I.I. Smolyaninov, Optics Letters, 19, 1601 (1994).

\bibitem{14} I.I. Smolyaninov, Phys.Rev.Letters 94, 057403 (2005).

\end{references}
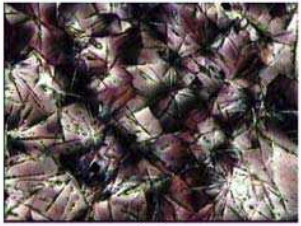
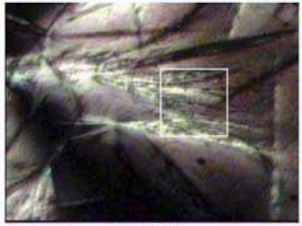
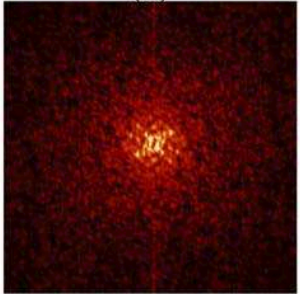
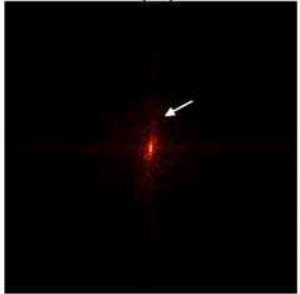
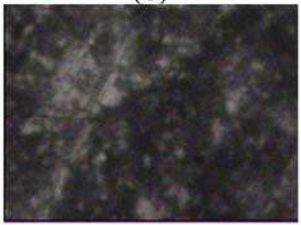
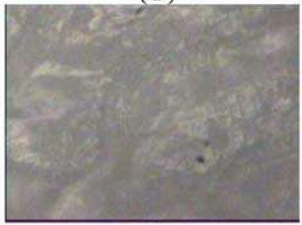
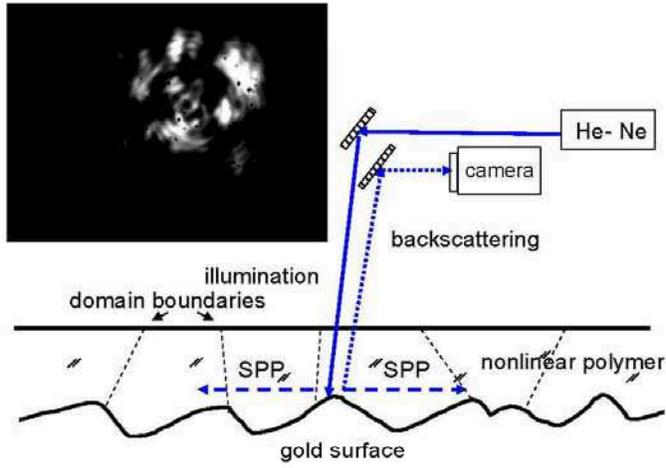
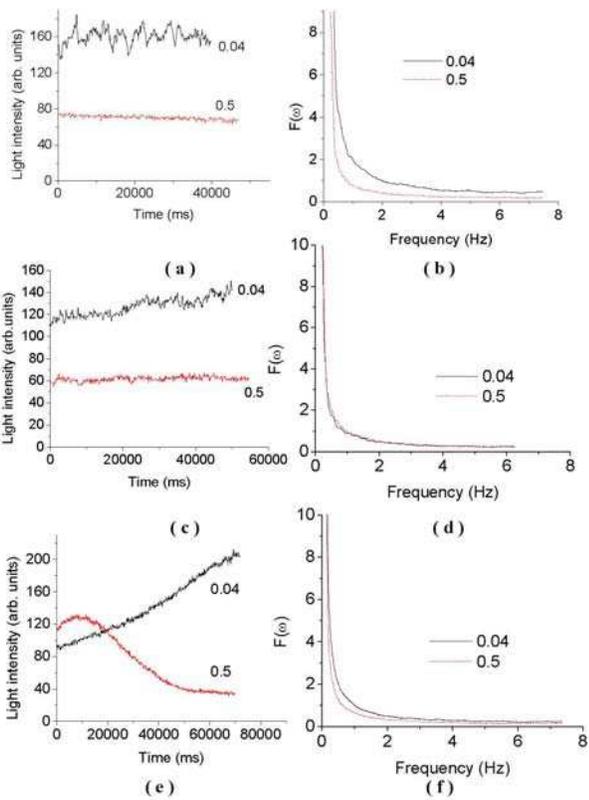
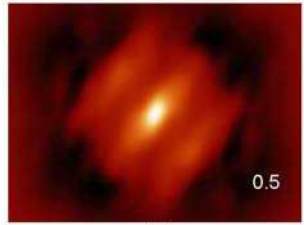
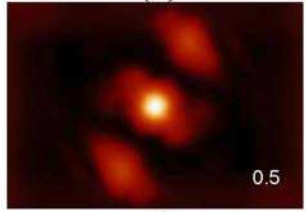
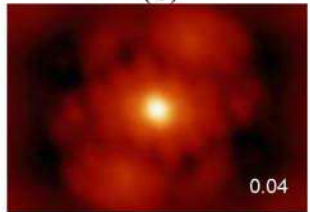
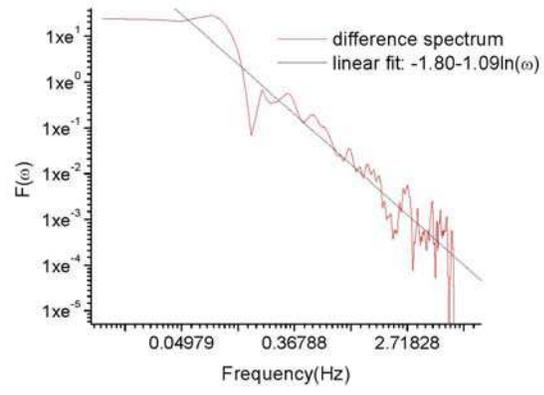
\end{document}